# 基于深度学习的代码生成方法研究进展*

杨泽洲[1], 陈思榕[1], 高翠芸[1], 李振昊[2], 李戈[3], 吕荣聪[4]

[1](哈尔滨工业大学(深圳) 计算机科学与技术学院, 广东 深圳 518055)
[2](华为技术有限公司, 广东 深圳 518129)
[3](北京大学 信息科学技术学院, 北京 100871)
[4](香港中文大学 计算机与工程系, 香港 沙田)
通讯作者: 高翠芸, E-mail: gaocuiyun@hit.edu.cn

**摘 要**: 本文关注根据自然语言描述生成相关代码片段的代码生成(Code Generation)任务. 在软件开发过程中, 开发人员常常会面临两种情形. 一种是通用功能的实现, 需要开发人员编写大量重复且技术含量较低的代码; 另一种是依赖于特定任务要求, 需要开发人员查询文档或使用其他工具才能完成的代码编写工作. 代码生成作为最直接辅助开发人员完成编码的工作受到学术界和工业界的广泛关注. 让机器理解用户需求, 自行完成程序编写也一直是软件工程领域重点关注的问题之一. 近年来, 随着深度学习在软件工程领域任务中的不断发展, 尤其是预训练模型的引入使得代码生成任务取得了十分优异的性能. 本文系统梳理了当前基于深度学习的代码生成相关工作, 并将目前基于深度学习的代码生成方法分为三类: 基于代码特征的方法、结合检索的方法以及结合后处理的方法. 第一类是指使用深度学习算法利用代码特征进行代码生成的方法, 第二类和第三类方法依托于第一类方法进行改进. 本文依次对每一类方法的已有研究成果进行了系统的梳理、分析与总结. 除此之外, 本文还汇总并分析了已有的代码生成工作中常用的语料库与评估方法, 以便于后续研究进行实验设计. 最后, 本文对代码生成方法研究进展进行总结, 并针对未来值得关注的研究方向进行展望.

# Deep Learning Based Code Generation Methods: Literature Review

YANG Ze-Zhou[1], CHEN Si-Rong[1], GAO Cui-Yun[1], LI Zhen-Hao[2], LI Ge[3], LYU Michael Rung-Tsong[4]

[1](School of Computer Science and Technology, Harbin Institute of Technology, Shenzhen, 518055, China)
[2](Huawei Technologies Co., Ltd., Shenzhen 518129, China)
[3](School of Electronics Engineering and Computer Science, Peking University, Beijing 100871, China)
[4](Department of Computer Science and Engineering, The Chinese University of Hong Kong, Hongkong, China)

**Abstract**: This paper focuses on Code Generation task that aims at generating relevant code fragments according to given natural language descriptions. In the process of software development, developers often encounter two scenarios. One is requested to write a large amount of repetitive and low-technical code for implementing common functionalities. The other is writing code that depends on specific task requirements, which may necessitate the use of external resources such as documentation or other tools. Therefore, code generation has received a lot of attention among academia and industry for assisting developers in coding. In fact, it has also been one of the key concerns in the field of software engineering to make machines understand users' requirements and write programs on their own. The recent development of deep learning techniques especially pre-training models make the code generation task achieve promising performance. In this paper, we systematically review the current work on deep learning-based code generation and classify the current deep learning-based code generation methods into three categories: methods based on code features, methods incorporated with retrieval, and methods incorporated with post-processing. The first category refers to the methods that use deep learning algorithms for code generation based on code features, and the second and third categories of methods improve the performance of the methods in the first category. In this paper, the existing research results of each category of methods are systematically reviewed, summarized and commented. Besides, the paper summarizes and analyzes the corpus and the popular evaluation metrics used in the existing code generation work. Finally, the paper summarizes the overall literature review and provides a prospect on future research directions worthy of attention.

## 1 引言

自计算机诞生以来, 虽然编程的形式随着硬件及软件的不断进步而不停迭代, 但是从事计算机技术行业的人员始终与编写代码的任务紧密联系在一起. 因此如何提高软件开发的效率和质量, 一直是软件工程领域的重要问题之一. 这一方面是由于在不同软件开发过程中往往存在大量相似代码复用的情况, 多次编写重复代码会大大降低开发人员的开发效率以及创造热情; 另一方面, 为完成特定任务相关而编写的代码往往是非重复的, 这样的代码编写工作为开发人员的编程能力带来挑战，开发人员往往需要联系代码上下文、查阅外部文档或借助其他工具以实现需求; 另外, 结构清晰、 功能完备的高质量代码能够使得软件开发过程明晰, 能够有效降低后期维护成本.

除了互联网领域的工作人员, 计算思维在当今信息社会对于每一位从业者都必不可少. 政府也在制定相关专门文件推动和规范编程教育发展, 帮助学生理解并建立计算思维[1]. 但事实上, 对于大多数没有经过系统化学习的人而言, 从零开始上手编程并完整完成一段能够实现具体功能的程序是极具挑战的. 编程作为一种完成人们所设想功能的手段, 本质上是一种工具, 但这样的学习门槛使得想法与实际操作之间存在差异, 在一定程度上阻碍了许多具有创意性思维程序的诞生.

程序自动生成方法是一项机器根据用户需求自动生成相应代码的技术. 智能化代码生成具有多元形式, 一般地, 根据实际应用场景以及生成过程所需要的信息, 可以将智能化代码生成分为代码生成(Code Generation)和代码补全(Code Completion)两个任务. 前者是指根据开发人员利用自然语言编写代码的需求(部分会加入输入输出样例、环境信息等), 机器生成特定编程语言的代码片段(部分方法加入后处理环节以保证代码的可执行); 后者则是指在开发人员编写代码过程中, 代码补全算法模型根据已编写代码上文自动理解开发人员的编写意图并补全代码. 根据补全的代码粒度, 又可以将其分为词元级别(token-level)以及行级别(line-level)[2]. 简单来说, 代码生成的输入是自然语言描述, 输出是能够在一定程度实现自然语言描述功能的代码片段; 而代码补全任务接收的输入是当前代码的上文, 输出的是当前代码的下文. 本文研究的智能化代码生成限定于前者, 即代码生成(Text-to-Code), 旨在根据自然语言描述生成特定编程语言的代码片段. 具体来说, 本任务关注软件实际开发过程中使用的高级编程语言, 如 C++, Java 和 Python 等.

传统的代码生成方法[3, 4]主要依赖于高质量的词汇表, 手工构建的模板和特定领域的语言特性, 要求程序员人工编写逻辑规则, 以便生成程序能够根据其设定规则生成符合逻辑的代码片段. 这种人工提取的方法具有很大的局限性, 不能够适应复杂多变的编程环境, 同时也增加了开发人员编写逻辑规则的开销. 因此随着人工智能和深度学习算法的进步, 自动提取词汇表及特征来生成可执行代码片段也得到长足发展.

事实上, 最近十年来, 利用机器学习和深度学习算法解决计算机各个领域相关问题的研究已成为一种趋势. 一些研究结果[5, 6]为将人工智能领域的相关算法模型应用到代码领域提供了理论基础, 即 AI(Artificial Intelligence) for SE(Software Engineer). 对于代码生成任务而言, 借助机器学习和深度学习算法, 利用数据驱动构建模型, 完成自动代码生成已成为程序自动代码生成任务的解决范式, 称为智能化代码生成. 智能化代码生成能够有效应对不同的开发环境, 提高软件开发的效率和质量, 减轻开发人员的压力, 降低代码编写的门槛.

然而, 代码生成问题在具体的研究过程中也面临诸多严峻挑战: 首先, 自然语言描述形式多种, 表达多

样，对于同样的函数描述实现可能一百个人就有一百种表达方式．因此，能否正确理解自然语言所描述的意图对于代码生成的质量具有决定性作用．其次，代码生成本质是生成类任务，使用到的模型在解码过程中往往伴随着巨大的解空间，针对复杂问题所生成的代码可能存在无法被执行或对于实际问题没有完全解决的情况，如何在其中找到正确的符合用户需求的代码仍需要被探索．因此，利用已有的外部知识库(例如 Stack Overflow 论坛)来提升代码生成模型的效果成为了一个可能的解决方案．上述这些问题虽随着模型的一步步增大带来的性能提升有所缓解，但仍未彻底解决．最后，目前对于代码生成的质量评估主要采用了自动评估的方式，评估的指标从机器翻译领域的指标转向基于测试样例的指标，这一定程度上有助于实际模型性能的评价．但对于实际场景中的代码生成，缺乏人工评价的板块，使得目前代码生成模型落地后的表现距开发人员的期待仍存在一定的差距．

## 2 研究框架

这里给出本文所研究的代码生成问题的数学定义．给定自然语言描述：$NL=\{nl_1,nl_2,\dots,nl_n\}$，智能化代码生成模型 $f_\theta$，目的是生成代码片段：$PL=\{pl_1,pl_2,\dots,pl_m\}$，有如下过程：$f_\theta(\{nl_1,nl_2,\dots,nl_n\})=\{pl_1,pl_2,\dots,pl_m\}$． 为了对智能化代码生成相关领域已有的研究工作和成果进行系统的梳理，本文使用 code generation、generating source code、generating program 和 program synthesis 作为关键词在 ACM Digital Library、IEEE Xplore Digital Library、 Elsevier ScienceDirect、 Springer Link Digital Library、 Google scholar 和 CNKI 在线数据库中进行检索．基于上述论文数据库中检索出来的相关文献，我们在人工筛选方法的基础上，通过分析论文的标题、关键词和摘要去除与代码生成无关的文献．接着我们递归地对每篇文献进行正向和反向滚雪球搜索[7]，最终选择出与主题直接相关的 54 篇高质量论文(截止到 2022 年 11 月).

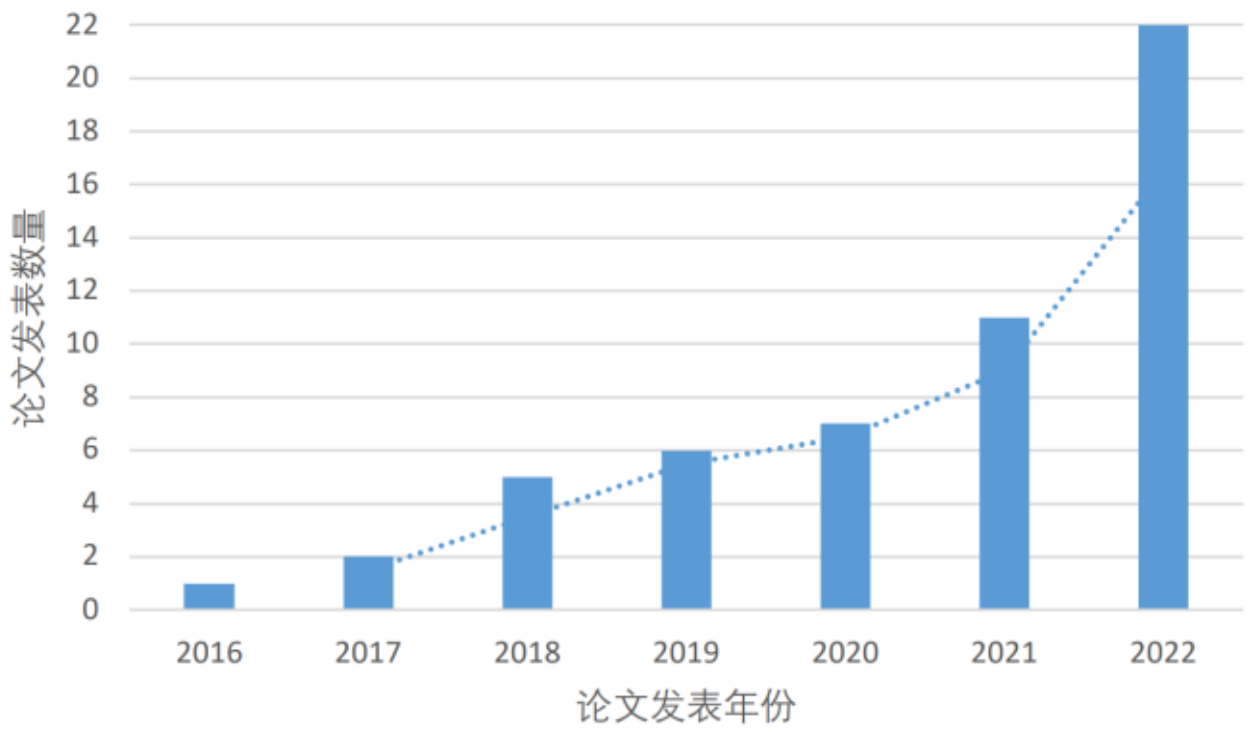

Fig. 1 Cumulative number of published papers for each year

图 1 相关论文每年累计发表的论文数

如图 1 所示，从 2015 年到 2019 年，智能化代码生成相关论文的数量呈上升趋势，说明智能化代码生成任务逐渐被学术界关注并取得了一些成果．2020 年预训练模型 CodeBERT[8]被提出，智能化代码领域掀起了利用大型语言模型在代码语料进行预训练，并在下游相关任务进行微调的热潮．因此 2021 年和 2022 年相关的论文大幅增加．尤其是 2022 年，论文数量达到了 22 篇，相关话题也在学术界和工业界之外引起了社会的广泛讨论．值得一提的是，企业的论文(如 PANGUCoder[9]与 AlphaCode[10])常常直接在网上进行模型架构的公开，为产品做技术支撑．虽然这些论文不会在期刊或会议上发表，但是其背后的思想及解决问题的工程方法具有很强的借鉴意义，不容忽视．另外，2022 年 12 月 ChatGPT[11]发布，引起了全社会的广泛讨论．虽然是一个对话模型，但 ChatGPT 被证实具有生成代码的能力．2023 年 3 月 Openai 发布的 GPT-4[12]模型同样在代码生成任务上取得了突出的表现，本文也将对此进行讨论．

Hu 等人曾在 2019 年对基于深度学习的程序生成与补全技术进行了系统完备的中文综述[13], 但是随着后续智能化代码生成任务相关研究成果的急剧增加, 领域内亟需对于此任务进行回顾与总结, 方便后续研究者在此领域继续深耕. 本文将主要内容放在生成高级程序语言所编写的代码生成任务上, 主要讨论基于深度学习(即神经网络架构)的相关模型. 根据代码生成方法的主要思想和核心部件, 本文主要将代码生成方法分为三类, 基于代码特征的代码生成方法、 结合检索的代码生成方法和结合后处理的代码生成方法, 后两者可以视作基于第一类方法进行改进的代码生成模型, 而第一类又可以根据使用的范式不同分为基于监督学习的代码生成方法和基于预训练的代码生成方法. 表 1 展示了本文所收集论文分类的概况.

本文主要关注智能化代码生成任务使用的模型以及算法的创新性, 算法的评估指标(及优劣), 使用数据集以及适用编程语言(即适用场景)四方面内容. 同时对之前工作中涉及到的数据集及评价指标单独进行了整理, 并对于不同数据集涉及的编程语言类型, 规模和相关论文使用的情况进行了统计. 除此之外, 本文还针对拥有检索增强模块以及后处理环节的论文进行单独讨论.

本文第 1 章为引言, 第 2 章介绍综述的整体研究框架. 第 3 - 5 章分别介绍基于代码特征、 结合检索以及基于后处理的代码生成方法的相关研究工作, 并进行讨论. 第 6 章汇总常用的代码生成数据集. 第 7 章介绍代码生成评估指标. 第 8 章对全文进行总结并对未来值得研究的方向进行了展望.

Table 1 Summary of existing studies of code generation based on deep learning

表 1 基于深度学习的代码生成现有工作的方法总结

| 方法 | | 相关文献 |
|---|---|---|
| 基于代码特征的代码生成方法 | 基于监督学习的代码生成方法 | Ling 等人[19], Yin 等人[20, 24, 25], Rabinovich 等人[21], Iyer 等人[22, 23], Sun 等人[26, 29], Wei 等人[27], Ye 等人[28] |
| | 基于预训练的代码生成方法 | Lu 等人[2], Feng 等人[8], Christopoulou 等人[9], Li 等人[10], Kanade 等人[30], Guo 等人[31, 43], Xu 等人[33], Clement 等人[38], Ahmad 等人[40], Wang 等人[41], Phan 等人[42], Chen 等人[44], CodeGeeX[50] |
| 结合检索的代码生成方法 | | Drain[52], Hayati 等人[53], Hashimoto 等人[54], Guo 等人[55], Xu 等人[56], Parvez 等人[58], Zhou 等人[60], Zan 等人[61] |
| 结合后处理的代码生成方法 | | Zhang 等人[59], Poesia 等人[62], Jain 等人[63], Wang 等人[64], Chen 等人[65], Le 等人[66] |

## 3 基于代码特征的代码生成方法

直观上, 代码生成任务可以简单抽象为机器翻译的问题: 将自然语言描述翻译为代码表示. 在机器翻译问题中最常用的模型是序列到序列模型(Squence2Sequence Model)[14], 其应用到代码生成任务的主要思想就是从训练数据中学习自然语言特征, 并利用代码特征进行生成. 在这个过程中需要用到大量的自然语言-代码对(<Text, Code>), 以便模型能够从训练数据中学习到双模态数据的对应关系, 因此本文将其称作基于监督学习的代码生成方法, 将在本章的第一小节进行介绍.

自 2017 年 Transformer 模型[15]的提出, 大型预训练模型在自然语言处理领域不断发展, 对计算机视觉甚至通用人工智能领域都产生了巨大的影响. 同样, 代码生成任务也积极引入大型预训练模型, 试图在已有挖掘代码特征的基础上结合预训练范式进一步提升模型性能. 预训练模型主要存在两个特征, 第一个是在预训练阶段其所需数据往往是无标注的数据, 通过还原掩藏掉的部分词元或片段来进行训练[16], 这样的训练过程被称为自监督训练. 第二个是模型的规模以及所需数据量都非常大, 这是因为许多工作[17, 18]都发现随着模型规模的增大, 数据量的增多和计算量的提升, 模型的性能就会不断提高. 与许多其他领域任务一样, 基于预训练的代码生成方法显著提高了代码生成的下限, 并逐渐成为最近几年研究代码生成问题的主要解决方案. 因此, 本文将基于预训练模型进行代码生成任务的代码生成方法单独列出, 作为本章的第二小节.

### 3.1 基于监督学习的代码生成方法

#### 3.1.1 简述

在基于监督学习的代码生成方法中，最常用的模型是序列到序列模型[13]，其对应的是编码器-解码器范式，主要包含编码器(Encoder)和解码器(Decoder)两部分，被广泛应用于自然语言处理中的生成类任务，同样也被代码生成类任务广泛使用．对于代码生成任务而言，编码器将输入自然语言描述转变为固定长度的向量表示，解码器则将编码器所产生的向量表示转变为符合所描述需求的程序输出，下面提到的相关模型均基于这样的框架实现并运行．与自然语言相比，代码自身具有规律化和模块化的特点，因此模型往往对解码的相关操作以及应用信息进行改进．在基于监督学习的代码生成方法中使用到的数据均为<Text, Code>的形式，目的是从这样的数据对中学习到文本和代码之间的对应关系，从而完成生成任务．

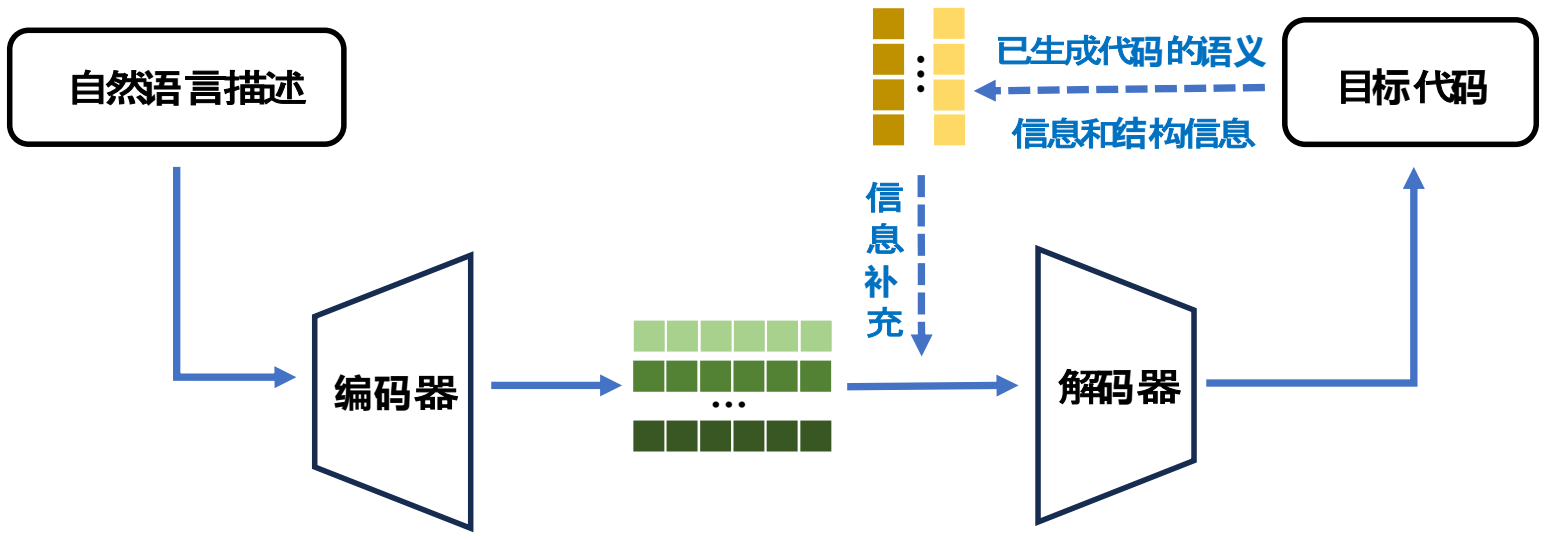

Fig. 2 The overview of the methods based on supervised learning

图 2 基于监督学习的代码生成方法概览图

图 2 展示了基于监督学习的代码生成方法概览，实线箭头绘制是使用传统的编码器-解码器架构实现代码生成的过程．首先使用编码器对自然语言描述进行编码，得到中间向量后输入到解码器以得到目标代码．除了直接使用编码器-解码器架构的模型完成代码生成任务，也有工作将代码的结构信息纳入模型之中．由于解码器解码的过程是自回归的，即上个时刻的输出作为这个时刻的输入．因此可以通过提取已生成代码的语义信息和结构信息来充实解码器的输入，图 2 中由虚线部分进行表示，从而解码生成更符合规范的代码．

#### 3.1.2 已有工作的分析

高级程序设计语言的智能化代码生成流程首先由 2016 年 Ling 等人[19]的工作定义．Ling 等人希望在卡牌游戏万智牌(Magic the Gathering)和炉石传说(Hearth Stone, HS)中，可以通过自然语言描述某一张卡牌的能力或效果，让模型自动地去生成对应的卡牌定义代码(高级程序语言，即 Java 和 Python)来减少开发人员编写卡牌效果的时间成本．作者利用序列到序列模型来实现自然语言到代码片段之间的转换．在传统注意力机制模型中，输入的注意力是由循环神经网络(Recurrent Neural Network, RNN)中每一个单元的输出经过计算加权得来．作者基于原有的注意力机制模型，在计算输入的注意力时考虑了每张卡牌自身的各项属性和功能对应的文本描述字段，并将不同的字段都用作计算注意力值．但是由于每一个输入字段的值域和输入规模不尽相同，因此作者利用线性投影层将不同输入投影到同一个维度和值域上．为了统一自然语言和代码之中的实体(如变量名等)信息，作者利用选择预测器(select predictor)来决定自然语言中需要保留的字段，并且结合序列到序列模型来生成代码的主体框架，最后再填入自然语言中的保留字段来实现对应卡牌代码的生成．

2017 年，Yin 等人[20]在 Ling 的工作上[19]做出了进一步改进，作者设计了一个能够将自然语言描述转换为目标编程语言的抽象语法树(Abstract Syntax Tree, AST)的模型，核心思想是显式地对目标语言的语法进行建模，并且将得到的语法信息作为先验知识融入模型训练之中，以此来限制模型的搜索空间，确保能够生成正确格式的代码．模型的编码器使用了双向长短期记忆网络(Bidirectional Long Short-Term Memory, BiLSTM)对自然语言进行编码，模型的解码器使用普通的长短期记忆网络(Long Short-Term Memory, LSTM)来构建抽象语

法树的序列生成过程. 在实际解码过程中, 解码器会结合之前已经生成对动作序列来影响最终抽象语法树的构建. 模型的目标是生成能够构建抽象语法树的动作序列而非代码. 根据所得的构建动作序列生成语法树, 并转化为相应的代码片段. 经过实验, 模型在 HS 数据集上达到了当时最好的结果.

同年, Rabinovich 等人[21]试图通过设计抽象语法网络(Abstract Syntax Networks, ASNs)作为标准编码器-解码器范式的扩展, 从而将代码的结构信息结合进代码生成的过程. 与常规的序列到序列模型输出代码不同, 抽象语法网络的输出为抽象语法树, 模型的编码器使用了双向长短期记忆网络, 模型的核心在于改变解码器, 使其具备与输出语法树结构平行的动态确定的模块化结构. 作者在文中使用了抽象语法描述框架(Abstract Syntax Description Language, ASDL)将代码片段表示为具有类型化节点的树. 模块化解码器本质为可以相互递归的模块集合, 其包含的四个模块根据给定输入分别以自顶向下的方式递归生成可执行代码的抽象语法树. 经过这样流程所产生树的语法结构就反映了生成过程的调用图, 即完成了代码生成任务. 作为一个模块化的编码器-解码器架构, 抽象语法网络适用于生成目标具有递归分解属性的情景. 在这样的场景下, 抽象语法网络可以提供与输出的固有结构非常相似的平凡的解码过程.

2018 年, Iyer 等人[22]在将自然语言映射到源代码的过程中加入了真实世界编程的上下文信息. 详细来说, 代码生成任务被进一步定义为在给定 Java 类环境信息的条件下, 利用文档生成成员函数. 同时, 作者提出的数据集 CONCODE 也被 CodeXGLUE[2]收录作为代码生成任务最常用的数据集之一. 在编码器端, 为了更好地利用输入中的 Java 类环境信息, 标识矩阵(Identifier Matrix)被用于对自然语言描述中的每一个单词进行编码; 为了对变量和方法进行更好地编码表示, 类型矩阵(Type Matrix)被用于标识方法名和变量名数据类型. 在解码器端, 作者利用两步注意力机制来计算自然语言描述中每一个 token 的注意力以及环境上下文中每一种类型和变量名的注意力, 最终利用第二步的注意力结果作为最后生成代码的先验信息. 除此之外, 由于模型生成的成员函数中有可能包含没有出现过的标识符, 所以作者利用监督复制机制来选择需要复制到代码中的标识符. Iyer 等人[23]在一年后提出了一种迭代方法, 通过反复整理, 从大型源代码语料库中提取最频繁深度-2 子树作为习语(idiom), 并训练模型能够在解码过程中应用这些习语. 简单来说, 这里的代码习语是指代码片段中最常见的结构(比如嵌套循环, 条件判断等), 蕴含丰富的代码结构信息. 相比于之前直接利用语法树解析代码结构的方法, 作者使用语法解析树中经常出现的深度为 2 的简化语法子树表示作为代码习语, 并将其直接应用到解码过程中, 该做法可以提升训练速度和性能, 使得模型更精准地输出对应的代码.

同样是在 2018 年, Yin 等人[24]则是在之前自己工作的基础上[20]提出了 TranX, 一个基于 transition 的代码生成模型. TranX 的核心在于 transition 系统. 结合一系列树构造动作, transition 系统将输入的自然语言描述转换为抽象语法树. 模型以抽象语法树作为中间表示, 对特定领域结构下的目标代码表示进行抽象. 最后再利用一个由神经网络参数化的概率模型对每一个假定的抽象语法树进行评分, 从而得到目标代码的抽象语法树, 进而完成代码生成的目的. 一年后, Yin 等人[25]在此基础上加以重排序(Reranking)技术进行优化, 主要使用了重排序模型对代码生成结果进行重排序来提升模型性能. 传统的生成模型往往使用近似推理(例如束搜索)来得到最好的 n 个候选输出, 重排序模型则是将根据自然语言描述对 n 个候选输出进行打分, 最终返回重排序得分最高的作为模型的最终输出. 详细来说, 重排序主要分为两个方面的内容, 一方面是根据生成的代码来重构真实输入文本的概率进行评估; 另一方面是利用匹配模型通过对齐, 比较和聚合三个步骤直接评估代码与输入文本的相关程度. 重排序模型也同样包含两个模块来实现上面所述两方面功能, 一是抽取生成重构特征(Generative Reconstruction Feature)的模块, 本质上是一个带注意力机制的序列到序列模型, 通过生成的代码极大似然估计出输入的文本; 二是计算判别匹配分数(Discriminative Matching Feature)的模块, 将生成代码与输入文本的 token 借助注意力机制的神经网络计算出结果. 除此之外, 作者还将 token 出现的频次作为一个特征纳入重排序的考虑范畴. 最后, 将上面三个特征进行加权求和纳入模型的推理过程就可以得到最终重排序的结果.

2019 年, Sun 等人[26]发现了程序代码比起自然语言描述包含有更多有意义的 token, 因此之前工作使用 RNN 来捕获长距离序列可能并不恰当. 作者在文中提出了基于语法的结构化卷积神经网络(Convolution Neural

Network, CNN), 通过预测抽象语法树的语法规则来完成代码生成任务. 详细来讲, 除了用来处理自然语言输入的卷积神经网络模块, 作者还设计了包含有基于树的卷积(Tree-Based CNN)和前序遍历的卷积(Pre-Order Traversal CNN)等多种卷积神经网络模块, 不同的模块可以通过结合注意力机制的池化层进行信息的聚合. 这样的做法本质上是在解码过程中进行了部分抽象语法树(即已生成的代码所构成的抽象语法树)信息的补充, 不同的卷积模块提取了抽象语法树中不同粒度的信息, 从而在聚合后达到增强模型的效果.

同年, Wei 等人[27]利用代码生成(Code Generation)任务和代码摘要(Code Summarization)任务两者的对偶性(duality)来同时提高两者的性能, 即代码生成模型和代码摘要模型被同步地训练. 作者提出了包含有代码生成模型, 代码摘要模型和对偶约束三部分的训练框架. 代码生成模型旨在将自然语言描述映射到源码片段, 代码摘要模型旨在将源码翻译成一段注释. 作者使用了相同结构的序列到序列模型来完成两个任务, 其中编码器使用了双向长短期记忆神经网络, 解码器使用了带有注意力机制的长短期记忆神经网络. 对偶约束是指在损失函数中添加正则项来约束两个模型的对偶性. 通过对偶任务的训练来提升模型的性能这样的思想也被运用到后续的研究当中[28].

2020 年, Sun 等人[29]使用 Transformer 架构来解决代码元素之间存在的长依赖问题, 并对模型进行修改提出 TreeGen, 使其能够结合代码的结构信息. 具体来说, TreeGen 共分为三个部分: NL Reader, AST Reader 和 Decoder. 其中, NL Reader 用于对自然语言描述进行编码, 本质上是多个块(block)的堆叠, 每个块都包含了三个不同的子层(自注意力, 门控机制和词卷积). AST Reader 用于对已生成的部分代码的语法树进行编码, 在编码过程中需要考虑包括预测规则和树结构的异构信息, 以便能够通过已生成的部分规则来预测接下来的生成规则. 具体来说, AST Reader 首先将代码表示为规则序列,然后使用注意力机制对规则进行编码, 最后使用树卷积层将每个节点的编码表示同它的祖先相结合. Decoder 则是用于聚合带有自然语言描述的生成代码的信息并预测接下来的语法规则, 作者在此使用了两个注意力层来结合 AST Reader 和 NL Reader 的输出, 再通过两个全连接层, 其中第一层包含一个 GELU 激活函数, 最终得到用于预测的抽取到的特征. 本文将基于 AST 的代码生成任务视作通过语法规则来扩展非终结符节点, 直至所有的叶子结点均为终结符停止的过程.

### 3.2 预训练模型

#### 3.2.1 简述

近些年来, 预训练模型在自然语言处理领域取得了巨大的成功. 从预训练模型架构来看, 可以分为编码器-解码器架构(Encoder-Decoder)、仅编码器架构(Encoder-only)和仅解码器架构(Decoder-only). 虽然模型总体架构并无形式创新, 但是其核心思想与之前的基于监督学习的模型不同. 预训练模型从大规模的无标注数据中通过自监督训练策略获取知识, 且绝大多基于 Transformer 架构进行设计. 首先在大规模的无标注数据集上对模型进行预训练, 然后利用预训练得到的表征在下游的有标注数据集上进行微调. 经过预训练的模型具有更好的泛化性, 微调后可以在多项任务上取得很好的结果. 类似地, 研究人员在代码领域也提出了对应的预训练模型, 并且在代码生成任务上取得了优异表现.

#### 3.2.2 已有工作的分析

CuBERT[30]是首个提出代码预训练模型的工作, CuBERT 继承 BERT[16]架构, 利用 6.6M 的 Python 代码语料进行训练. 预训练的 CuBERT 模型在五个分类任务上优于基线模型 LSTM. CodeBERT[8]是第一个多编程语言的大型双模态(指自然语言描述 NL 与编程语言 PL)预训练模型, 后续在多个下游任务上被广泛使用. 用于预训练的编程语言包括: Ruby、JavaScript、Java、Python、Go 和 PHP. CodeBERT 的预训练任务为掩藏语言建模(Mask Language Modeling, MLM)[16], 预训练通过随机掩藏模型中的某些词再让模型去预测被掩藏的词来完成, 有助于提高模型对于无标注数据的理解能力. 在实际场景当中, CodeBERT 被用来作为编码器对输入的文本或代码进行编码, 然后应用到各式各样的下游任务中, 例如克隆检测, 漏洞检测以及代码检索等.

由于 CodeBERT 仅仅包含文本的语义信息, 所以 Guo 等人[31]在 2021 年提出将代码的结构信息数据流纳入预训练的过程之中并提出 GraphCodeBERT. 作者对 MLM 进行了改进: 不仅掩藏文本信息中的一些单词, 而且会在代码的数据流图中随机掩藏某些数据节点然后让模型去预测. 实验证明在预训练过程中显式地去考虑

代码的结构信息可以极大地提高代码对模型的理解能力并提高在下游任务当中的性能.

上述代码预训练模型仅包含编码器端, 这种架构的预训练模型在理解任务上的效果较好, 但无法很好地完成生成式任务. 为了更好地完成根据自然语言描述进行代码生成(Text-to-Code)的任务, CodeXGLUE[2]中提出了 CodeGPT 模型, 这是一个由代码语料进行训练, 与 GPT-2[32]完全同架构的 12 层 Transformer 解码器模型. 与仅编码器架构的模型相比, 仅解码器架构能够更好地完成生成代码的任务.PolyCoder [33]同样采用 GPT-2 的架构, 使用 12 种编程语言的 249GB 数据进行预训练. 在代码生成任务上, 大多数模型主要基于 Java 和 Python 语言进行预训练和代码生成. 而 PolyCoder 的预训练数据中 C 语言的数据量占比最大, 因此在 C 语言的代码生成上取得了更好的表现.

Nijkamp 等人[34]开源了大型预训练模型 CodeGEN, 模型参数高达 16.1B, 依次在 THEPILE, BIGQUERY, BIGPYTHON 三个数据集上进行训练, 数据量超过 800G. 与之前直接将自然语言输入给预训练模型不同, 作者提出利用大型预训练模型进行对话式程序生成的方法: 作者将代码生成的过程描述为用户和系统之间的多轮对话. 用户分多次为代码生成系统提供自然语言并接收系统的反馈, 这样用户与系统一起在多轮对话后完成代码生成. 分步提供自然语言的方式可以将较长且复杂的意图分解为多个简单的意图, 减少每一轮对话中模型的搜索空间. 除此之外, 作者还开发了一个多轮编程基准来衡量模型的多轮编程能力. 实验结果表明, 与单轮提供输入相比, 以多轮方式提供意图显着改进了代码生成的性能, 验证了对话式代码生成范式的有效性.

虽然在大规模无标注的代码数据上进行预训练, 再通过代码生成数据集对预训练模型进行微调在一定程度上取得了不错的效果, 但是为了训练一个性能优异的代码生成模型, 大量高质量的<Text, Code>数据必不可少, 然而构建这样的数据集的时间和经济代价都十分高昂. Zan 等人[35]试图通过提出 CERT(for sketCher and gEneRaTor), 一种基于模板(sketches)的持续预训练方法来解决面向调库的代码生成问题. 与单独的代码片段相比, 面向调库的代码片段往往拥有更加相似的代码结构和组织方式. 因此, 作者将面向调库的代码生成任务拆分为两个子任务: 一是将这样代码中用户自定义的变量和常量进行匿名化处理得到代码模板; 二是根据代码模板填充细节, 从而完成代码生成任务. 作者根据两个子任务将模型分为 Sketcher 和 Generator 两个模块. Sketcher 可以根据输入的文本生成多个候选的代码模板, 并选择出现最多次数的代码模板作为其输出. Generator 接收来自 Sketcher 输出的代码模板以及模型本身的输入文本, 生成最终的代码. Skecher 和 Generator 使用模型相同, 这里作者使用了自己训练的 PyCodeGPT 作为 Sketcher 和 Generator 的基础模型. PyCodeGPT 采用了 GPT-Neo[36]的模型架构, 在作者从 Github 上收集的 96GB Python 数据上进行预训练, 以此来达到接近中等规模 CodeX 效果的目的. 实验结果表明, CERT 通过让 Sketcher 和 Generator 在无标注的数据集上持续训练的方法, 取得了面向调库的代码生成任务上性能的提升.

之前的代码生成模型都是从左到右生成代码序列, 然而在实际开发过程中, 代码很少直接以从左到右的方式编写, 而是完成部分代码编写后反复编辑和完善. InCoder[37]打破了先前从左至右的代码生成预训练模型范式. 这是一种统一的生成模型, 可以执行程序合成(从左到右生成)以及编辑(通过掩藏和填充). InCoder 的模型架构继承自 GPT 架构, 不同点在于其对训练语料进行顺序打乱预测. 该方法随机选择一个跨度并将其替换为掩码 token, 并将跨度放置在序列之后作为目标. 利用这样处理后的语料进行训练, 模型就具有填充双向上下文的能力. 这样 InCoder 不仅可以从左到右预测 tokens, 而且可以根据两端的 tokens 预测中间的 tokens, 实现了填充式的代码生成技术. 这是第一个能够填充任意代码区域的大型生成代码模型, 这种以双向上下文为条件的能力大大提高了代码生成任务的性能.

除了单独使用 Transformer 的编码器或解码器结构,也有相关工作同时使用 Transformer 的两端.

Clement 等人[38]基于 T5(text-to-text transfer transformer)[39]提出了多模态的翻译模型 PYMT5. 该模型可以同时完成代码生成和代码摘要任务. PYMT5 使用相似子序列掩藏目标(similar span-masking objective)任务进行预训练. 子序列掩藏目标是指随机采样一些连续的 3 个 token 的子序列并使用特殊标记(例如[MASK 0])对其进行掩藏, 然后训练序列到序列模型来补全这些掩藏的 token, 训练目标包含了被隐藏的 token 及其序号. PYMT5 的性能优于在相同数据集上进行预训练的 GPT-2[32].

Ahmad 等人[40]提出的 PLBART 也是一个编码器-解码器模型，在预训练过程中，与 CodeBERT 做法一样，随机掩藏某些单词，但是 PLBART 输出的是一个完整的文本或单词，其中包括了被掩藏的单词. 通过这种训练方式，作为一个序列到序列模型的 PLBART 就可以在预训练阶段在编码器和解码器端同时学习到较好的初始化点，让预训练好的模型可以更快更好地应用到下游任务当中，从而提高了模型的代码生成能力.

更进一步地, Wang 等人[41]在 2021 年提出了 CodeT5. CodeT5 在预训练阶段充分考虑代码特点. 作者从代码片段中抽取标识符，并重点让模型去预测这些在代码中具有实际意义的单词，实验证明 CodeT5 在包括生成任务在内的多项软件工程领域任务均取得了更好的效果. 同一时期, Phan 等人提出的 CoTexT[42]同样使用了编码器-解码器架构，模型初始化也使用了 T5. 最大的区别在于为了缩小预训练和调优之间的差异, CodeT5 将代码生成与代码摘要视作对偶任务(dual task)，并利用<Text, Code>双模态数据训练模型完成一个双向生成的转换. 对比两篇文章在代码生成任务上的实验结果, CodeT5 的整体效果优于 CoTexT. 但是如果不利用双模态数据的对偶性, CodeT5 和 CoTexT 效果相当. 这样的实验结果侧面证明了将代码生成和代码摘要任务作为对偶任务能够提升模型的生成能力.

Guo 等人[43]提出了一个统一的代码预训练模型 UniXcoder，可以同时兼容编码器-解码器架构、仅编码器架构和仅解码器架构. 该模型使用带有前缀适配器的掩码注意力矩阵(mask attention matrices with prefix adapters)来控制模型的行为. 在下游任务使用该模型时，只需在输入时显式地加上前缀<encoder-only>、<decoder-only>或<encoder-decoder>就可以使用对应的模型架构. 此外，为了学习代码的语义嵌入，作者提出了两个新的预训练任务：多模态对比学习(multi-modal contrastive learning, MCL)和跨模态生成(cross-modal generation, CMG). 这两个任务分别利用了 AST 和代码注释的信息，实验证明二者都可以增强 UniXcoder 对于代码的理解能力，并提升模型在下游任务上的性能.

越来越多基于 Transformer 体系结构的大型预训练模型被提出并在代码生成任务上取得突出结果. 因此，一些企业、机构着手于将大型预训练代码生成模型落地，试图为广大开发人员提供便利，并在此过程中为业界提供了大量优质的代码生成模型.

2021 年末, OpenAI 最早发布的 CodeX[44]. 这是一个基于 GPT-3 并在公开数据集上预训练得到的大规模模型，基于 CodeX 的 Copolit[45]插件也已成为代码生成辅助工具的标杆. 在 CodeX 论文中提出的 HumanEval 数据集也成为后续代码生成的常用基准数据集之一.

2022 年初, DeepMind 公司研发的展现出强大编程能力的 AlphaCode[10]在新闻上号称打败了一半的程序员. AlphaCode 本质上也是基于公开代码仓库进行预训练的大规模模型. 与 CodeX 不同, AlphaCode 更专注于竞赛题目的编写，因此选用了完整的 Transformer 架构的模型，便于更好地理解较长的由自然语言描述的题目，同时在调优时也选择了 CodeForces[46]的竞赛题目.

国内，华为推出的 PanGu-Coder[9]基于 PanGu-alpha[47]模型在公开代码数据集上进行预训练. 之后基于 PanGu-Coder 开发的 CodeArts 插件[48]也已在实际开发场景中拥有不错的表现，对标基于 CodeX 的 Copoilt.

2022 年, aiXcoder[49]团队陆续推出了用于 Java 代码补全的 13 亿参数量的 aiXcoder L 和 130 亿参数量的 aiXcoder XL 服务. aiXcoder L 基于 GPT-2, 在开源 Java 代码上训练得到. aiXcoder XL 基于自研的 masked language model 框架，能做到单行、多行以及函数级代码补全和生成.

清华大学联合鹏城实验室共同推出的大规模代码生成预训练模型 CodeGeeX[50]，采用了标准的 Transformer 架构，在公开代码仓库中的超过 20 多种编程语言上进行预训练，能够支持高精度的代码生成，同时支持代码片段在不同编程语言间进行自动翻译转换.

表 2 汇总了本节提及的代码生成预训练模型.

除以上专门用于代码任务的相关模型, ChatGPT[11]虽作为问答模型，但也被证实具有代码编写的能力，自 2022 年 12 月发布以来就引起了学术界和工业界的广泛讨论. ChatGPT 能够适应不同的问题情境给出接近甚至超过人类的回答，并具有一定推理和代码编写能力. ChatGPT 与 InstructGPT[51]使用了相似的训练方式. 主要流程分为三个步骤，第一步是搜集带标记的数据，使用监督学习策略对已有的 GPT 模型进行调优；第二步是搜

Table 2 Statistics of code generation pretraining model

表 2 代码生成预训练模型概况

| 预训练模型 | backbone | 模型规模 | 预训练数据集 | 数据量 | 训练语言 |
|---|---|---|---|---|---|
| CuBERT | BERT | - | Python from Github | 7.7 million unique files | Python |
| CodeBERT | BERT | 125M | CodeSearchNet | 20GB | Ruby/JavaScript/Java/ Python/Go/PHP/English |
| GraphCodeBERT | BERT | 125M | CodeSearchNet | 20GB | Ruby/JavaScript/Java/ Python/Go/PHP/English |
| CodeGPT | GPT-2 | 124M | Python and Java from CodeSearchNet | 1.1million Python functions and 1.6 million Java functions | Java/Python |
| CoTexT | T5 | - | CodeSearchNet AND Java and Python from BigQuery | - | Java/Python |
| CodeT5 | T5 | 60M/223M/ 770M | CodeSearchNet and C/C# datasets | 8.35G | Ruby/JavaScript/Go/ Python/Java/PHP/C/C# |
| PLBART | BART | 140M | Java and Python from BigQuery AND SO posts | 655G | Java/Python/English |
| UniXcoder | - | - | CodeSearchNet | - | Ruby/Java/Python/PHP/ Go/JavaScript |
| PyCodeGPT | GPT-Neo | 110M | high-quality python files | 96GB | Python |
| PolyCoder | GPT-2 | 2.7B | a mixture of repositories from GitHub | 249GB | C/C++/C#/Java/JavaScript/ Go/PHP/Python/Ruby/ Rust/Scala/TypeScript |
| CodeGen | - | 350M/2.7B/ 6.1B/16.1B | THEPILE/BIGQUERY/ BIGPYTHON | THEPILE 825G | C/C++/Go/Java/ JavaScript/Python/English |
| InCoder | - | 1.3B/6.7B | content from StackOverflow | 159G | PTYHON and 28 other languages |
| CodeX | GPT-3 | 300M/2.5B/ 12B | Python from GitHub | 159G | Python |
| AlphaCode | - | 300M/1B/ 3B/9B/41B | a snapshot of github | 715.1G | C++/C#/Java/JavaScript/ Lua/PHP/Python/Ruby/ Go/Rust/Scala/TypeScript |
| PanGu-Coder | PanGu-alpha | 317M/2.6B | Python from GitHub | 147G | Python |
| CodeGeeX | - | 13B | open-sourced code datasets, The Pile and CodeParrot | - | C++/Python/C/Java/ JavaScript/Go/HTML/PHP/ Shell/CSS/Others |
| aiXcoder L | GPT-2 | 1.3B | Java from GitHub | - | Java |
| aiXcoder XL | - | 13B | Open-sourced code from GitHub | - | Java |

集来自人类反馈的比较数据，从而训练一个分类器(称为 reward model). 该分类器用于评估 GPT 模型生成若干答案的质量；第三步是借助第二步训练好的 reward model，利用强化学习策略对模型进行进一步优化. 原有的提示学习范式是通过调整模型的输入使得下游任务更好地适配模型，而通过 InstructGPT 的训练过程及 ChatGPT 的出色表现可以发现，模型的性能尚未被充分挖掘，且可以通过人为标注反馈的方式，让模型更好地去理解用户的意图，达到让模型适配用户的目的.

OpenAI 在 2023 年 3 月发布 GPT-4[12]，在 GPT 系列模型的研究道路上更进一步，利用更多的数据和更庞大的计算资源来构建一个更为强大的大型语言模型. 其拥有比 ChatGPT 更加优秀的推理能力和上下文学习能

力. 实验结果发现, GPT-4 进一步提升了零样本场景下的生成能力, 在包含代码生成任务在内的多项人工智能任务上都能够实现最优表现.

### 3.3 小结

本章主要对基于代码特征的代码生成方法相关工作进行了叙述与梳理, 并将监督学习与预训练-调优范式分为两节进行概述.

在基于监督学习的代码生成方法中, 主要依托于编码器-解码器架构进行自然语言描述与代码特征的挖掘, 试图从训练数据中学习到自然语言与代码之间的对应关系. 除此之外, 一些工作针对解码器生成代码的部分进行了改动, 试图将更多代码相关的规则性信息融入其中, 取得了比仅使用 token 进行代码生成更好地性能.

在基于预训练的代码生成方法中, 主要对引入代码生成任务的预训练模型进行介绍. 这一小节的方法的骨干模型(backbone)大多来自于自然语言处理领域的大型语言预训练模型. 将其应用于代码生成任务的一个大致流程为: 收集大量无标注的代码语料, 其中包含代码注释, 利用试图还原掩藏掉的部分代码片段对模型进行预训练, 最后再使用<Text, Code>数据对预训练模型进行调优.

最后, 本章节讨论了 ChatGPT 和 GPT-4 在代码生成任务上的应用. ChatGPT 和 GPT-4 所具备的强大语言理解能力和一定程度的推理能力引起学术界和工业界的广泛关注, 其背后通过反馈调整使模型来匹配用户的思想, 为进一步挖掘大规模语言预训练模型提供了新的解决思路.

## 4 结合检索的代码生成模型

### 4.1 简述

对于第三章中的代码生成模型而言, 其代码生成过程的解空间过大, 这种现象在预训练模型中的表现愈发明显[44]. 互联网上存在的代码片段数量非常庞大, 对于用户的绝大部分要求而言, 其他开发人员可能存在过类似的需求, 并已经与他人协作或自行完成相关代码的设计与编写. 因此, 检索出类似已存在的代码模板这个任务本身就是有利于用户更深刻理解任务的.

编写代码是一个开放域(open-domain)的问题, 即在编写代码过程中不可避免需要参考前人工作与他人的编程思路, 而在第二章中介绍的大部分工作则是直接将其视作封闭性质的任务(closed-book)[52], 即只是根据训练数据的模式来完成代码生成的任务. 从这个角度来看, 引入外部知识库对原有训练的模型进行补充是有意义有价值的. 事实上, 利用自然语言描述的编码作为解码过程中的先验可能是不足的. 为了让已有代码生成模型与外部代码知识数据库相结合, 有相关工作利用检索操作对代码生成过程进行增强. 通过检索相似代码帮助解码器进行代码生成, 以减小解码空间, 最终提升生成代码的质量. 并且, 在实际开发过程中, 面对不熟悉的编码任务, 开发人员的效率会大打折扣. 针对这种情况, 结合检索的代码生成由于更符合大多数开发人员习惯而具有天然的优势, 因为模型可以通过检索获取外部知识, 弥补模型的信息缺陷. 因此, 本章节对结合检索的代码生成模型进行讨论.

### 4.2 已有工作的分析

Hayati 等人[53]首次将检索技术引入代码生成任务, 提出 RECODE 模型, 针对 Yin 等人[20]提出的模型不能正确生成复杂代码的问题进行改进. 文中使用 TranX[24]中 transition 的构造规则来创建抽象语法树. 具体来说, RECODE 首先从训练集中检索与输入句子最相似的自然语言描述; 然后根据对应的代码片段抽取语法树的生成行为序列. 在推理过程使用启发式方法, 根据前面得到的生成行为序列来改变最终解码过程中语法树构建特定行为的概率, 从而提升模型生成代码的性能.

Hashimoto 等人[54]同样在代码智能化任务中使用到了检索技术. 作者认为编辑并更改检索得到的代码比直接生成代码更简单有效, 并基于此想法设计了一种检索并编辑的框架(retrieve-and-edit framework). 该框架下的模型需要首先根据自然语言描述的输入与训练集中的自然语言进行相似度计算, 并检索到相关的代码,

然后再编辑检索到的代码使其能够更加符合自然语言描述的任务.此框架包含检索器(Retriever)和编辑器(Editor)两部分内容. 检索器是一个序列到序列模型. 首先利用训练数据对模型进行训练, 目的是将自然语言输入映射到能够重构为代码输出的向量; 训练完成后利用检索器中的编码器对输入的自然语言进行编码得到向量表示, 与训练集中其他自然语言描述计算相似度得到检索结果. 对于编辑器, 作者使用了一个标准的带有注意力和复制机制的序列到序列模型完成对代码的编辑.

Guo 等人将代码检索与元学习技术相结合, 用来提高模型代码生成的能力[55]. 作者提出了一种上下文感知的编码器-解码器模型来作为检索器, 将上下文环境(文中是指 Java 中的类环境信息)以及自然语言描述输入编码作为隐变量用于检索; 此外将检索出的数据点利用模型无关元学习(model agnostic meta-learning, MAML)训练范式快速地将检索数据适应到对应的任务之中. 作者沿用了先前工作[54]中的检索器, 本质上是利用编码器对上下文以及自然语言进行编码完成检索. 在实际训练过程中, 先从训练数据中采样一部分作为测试集, 使用检索器对除测试集之外的数据进行检索, 得到训练集, 以此来定义一个任务. 通过不同任务对模型进行的参数更新及选择, 这就是代码检索与元学习技术相结合的训练过程. 经过这种训练方式, 模型可以快速适应到检索器所提供的数据中. 作者在两种范式上均进行了实验, 分别是 MAML 和检索并编辑(即上文提到的 retrieve-and-edit framework[53]), 结果表明 MAML 可以更精准地生成代码.

Xu 等人[56]针对自然语言注释与代码数据的缺乏的问题, 基于 TranX 模型[24]引入了两个外部知识库来进行数据增强, 帮助模型更好地进行预训练, 并提升模型的性能. 作者使用的两个外部知识库分别是从 Stack Overflow 上和 API Documentation 上提取出来的<Text, Code>数据对. 考虑到 API Documentation 上处理得到的数据虽然没有噪音但是可能与开发人员的实际开发情境不符, 作者根据 CoNaLa[57]以及直接爬取 Stack Overflow 的数据(同样采用了 Yin[57]的方法)构建了真实用户产生的<Text, Code>分布, 并对 API Documentation 得来的数据进行重采样以消除文档与实际用户使用情况之间的差异. 这种方法与模型无关, 作者首先引入外部知识库的数据对模型进行预训练, 然后在人工注释的数据集 CoNaLa 上进行调优, 使得模型能够很好地完成代码生成任务.

Parvez 等人[58]认为之前结合检索的生成模型只是借助高质量的代码和文档信息对模型进行训练, 而在生成过程中并没有显式地利用这些额外信息. 因此作者提出 REDCODER(Retrieval augmentED CODe gEneration and summaRization)框架, 利用检索对生成模型的输入进行增强, 以此来达到最大程度利用额外信息的效果. 同时, 检索增强的方法是双向的. 一方面可以使用检索到的代码来完成代码生成任务的增强, 另一方面也可以通过检索到的文本信息对代码摘要任务进行增强. 详细来讲, REDCODER 主要分为两部分, 第一部分是检索器(Retriever), 其中包含两个编码器, 分别对代码和自然语言描述进行编码, 根据编码后的向量相似度检索出相关的源码(用于代码生成)或摘要(用于代码摘要); 第二部分为生成器(Generator), 将检索到的源码/文本内容与原始的模型的输入相结合来产生目标输出, 通过检索增强输入从而提升了 REDCODER 的代码生成能力. 对于代码生成任务而言, 通过交换输入输出可以得到代码摘要任务, 反之亦然, 这样的任务被称为对偶任务(dual task), 其两个任务的输入输出及模型示意如图 3 所示. 利用对偶学习进行模型增强的过程可以概括如下: 使用两个相同的基础模型完成代码摘要和代码生成. 两个模型仅有输入输出不同, 采用联合训练的方式完成模型性能的提升. 将代码生成任务与代码摘要任务视作对偶任务在其他工作中也有涉及, 用以提升代码生成任务的性能[27, 28, 58, 59].

Drain 等人[52]对于提供的 docstring(即描述需求的自然语言)与数据库中的文档分别创建编码器, 并针对 CodeSeachNet 问题中提出的 NBoW 算法进行了改进. NBoW 算法在使用 BPE 分词后直接对序列进行编码. 序列表示取自词元平均(mean-pooling), 最大(max-pooling)或基于注意力的加权和(attention-like weighted sum)三者之一. 作者在本文中则是直接将序列表示的三个来源直接进行线性组合, 称为混合表示, 最后再通过相似度计算得到前 K 个样板方法的代码. 该工作通过实验比较证明了基于这样混合表示的检索算法要优于之前的 ElasticSearch, NBoW 以及 DPR(Dense Passage Retrieval)三种常用的检索算法. 为了得到更高的信息密度, 检索到更多有意义的<Text, Code>, 作者没有直接使用 Stack Overflow 上问答的原始数据, 而是使用 CoNaLa[56]作

为 Stack Overflow 的替代数据集进行检索. 作者在文中使用的生成模型是 BART-large, 并在掩藏过的 Python 代码和 Stack Overflow 问答对的混合数据上继续训练. 在实际生成过程中, 生成模型将文本描述, 函数声明以

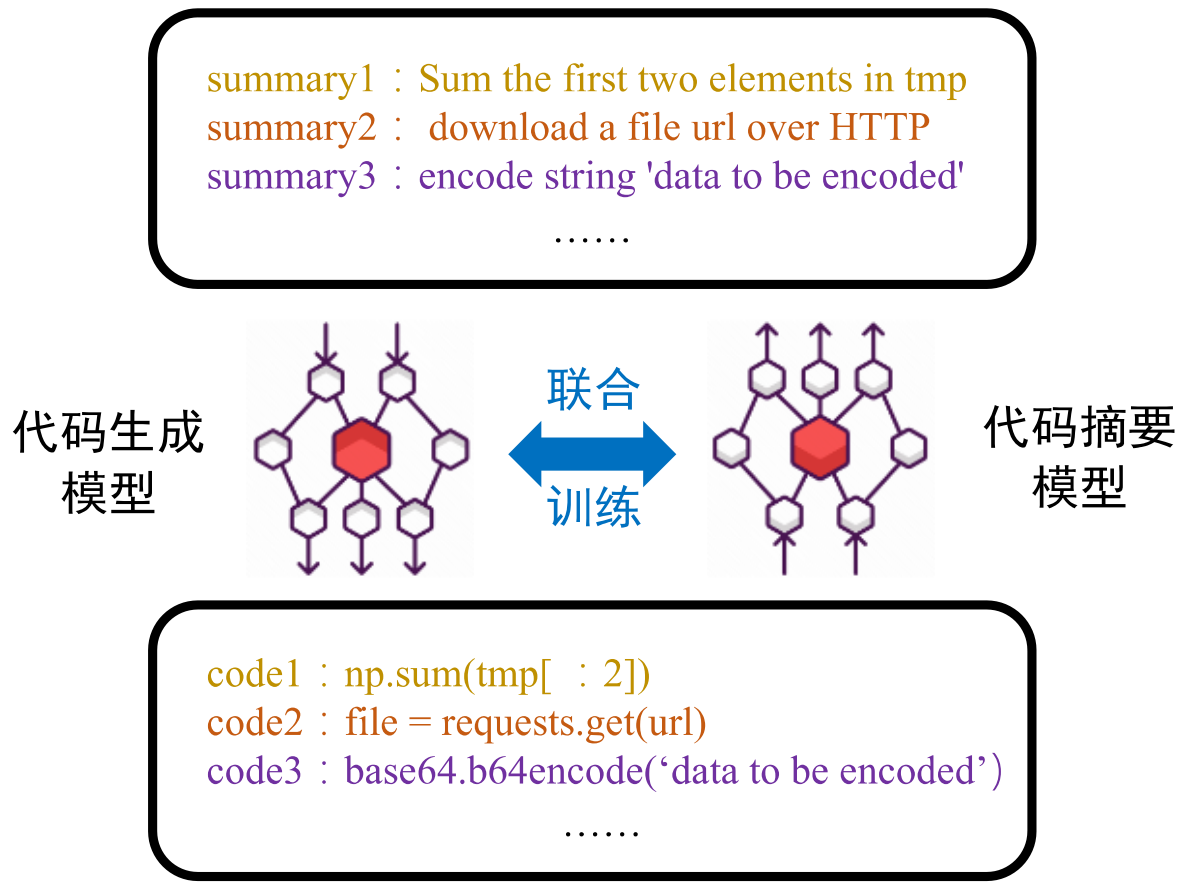

Fig. 3 Input and output of the dual task including code generation and code summarization

图 3 代码生成与代码摘要任务作为对偶任务的输入与输出

及检索到的前 K 个样板方法的代码作为输入, 最终得到所需方法具体实现的代码.

上述基于检索的工作都是在训练集中检索相似数据, 这些方法假定所有库和函数调用都在训练数据中存在, 因此在本质上无法借助训练数据中没有的代码对模型进行增强, 同时, 希望生成的代码可能会存在调用模型未知的函数和库的情况, 如果只是从训练数据中进行检索, 模型是不具备泛化能力的. 实际上, 训练数据不可能包含所有的库. 随着计算机科学的发展, 公开可用的 API 库也在不断增长变化, 模型不可能通过仅在现有代码库上训练就实现与所有可用的 API 保持同步. DocPrompting[60]同样使用检索的方式进行代码生成, 不同的是, 其维护一个文档库, 该文档库是独立于模型的外部数据存储, 并且允许进行频繁的更新. 无需重新训练模型, 模型可检索并利用动态更新的文档库, 从而生成模型训练时从未见过的代码. 实验表明, 这种检索的方式可以有效提升代码生成的性能. 同时, 作者在文中提到的对外部文档库进行更新在一定程度上能够缓代码生成作为开放域任务的普遍问题.

在实际开发过程中, 公司或团队内部的软件开发者共享工程代码与文件是普遍的. 考虑到安全性, 这些代码往往不会公开, 而是作为私有仓库在公司内部进行共享流转与使用. 由于私有仓库中的代码往往不会作为训练数据用于模型的训练, 因此希望针对私有仓库生成代码对于已有模型是极具挑战的. 为解决这个问题, Zan 等人[61]设计了一个框架使得预训练模型拥有生成使用私有仓库代码的能力. 受到开发人员学习使用私有库进行代码编写过程的启发, 作者试图通过构建检索并生成的框架来模拟学习私有 API 文档以及调用所学 API 来完成所需功能的过程. 作者在框架中创建了两个模块来完成这两项功能. 第一个模块是 APIRetriever, 其作用是基于编程问题和 API 文档来检索到有用的 API, 并设计了方便的交互方式来了解用户的需求, 本质是将自然语言描述与 API 信息利用 BERT 进行编码并进行相似度计算. 第二个模块是 APICoder, 其作用是直接使用已有的语言模型去调用私有的 API 完成描述的任务. 为了能够让语言模型知道怎样更好地调用 API, 作者针对 APIRetriever 检索到的代码文件进行分块, 并为每个代码块设置相应的提示(prompt), 基于 CodeGen 继续在包含了公开库 API 信息的代码库上进行训练, 称为 CodeGenAPI.

### 4.3 小结

本章主要对结合检索的代码生成方法相关工作进行了简单介绍. 本章中提到的方法并没有设计新的模型架构, 而是通过结合检索的方式生成代码片段. 如图 4 所示,结合检索的代码生成方法大致可以分为两种形式, 第一种如图 4 (a) 所示: 从训练数据或是外部知识库检索出与目标代码相关或相似的代码片段, 将其作为模板进行编辑并返回结果; 第二种如图 4 (b) 所示: 将检索得来的结果作为原有模型的输入, 试图通过为自然语言描述补充代码信息的方式提升模型的性能. 结合检索的代码生成方法能够降低生成模型解空间的规模, 同时也能在一定程度上利用外部知识库使模型跳脱出训练时学习到的固有规则, 有助于提高生成代码的多样性和可靠性.

在上述检索的过程中, 大多使用稠密检索方法对代码进行检索, 就是将自然语言描述作为查询的键(key), 并经模型处理后借助向量进行表示, 在训练数据或外部数据库查找与查询键相关的代码片段进行返回. 在这个过程中, 模型向量表示的能力决定了检索结果的质量. 因此, 尝试多种自然语言表示形式与不同的模型表征方法可能是检索增强的一条未来之路.

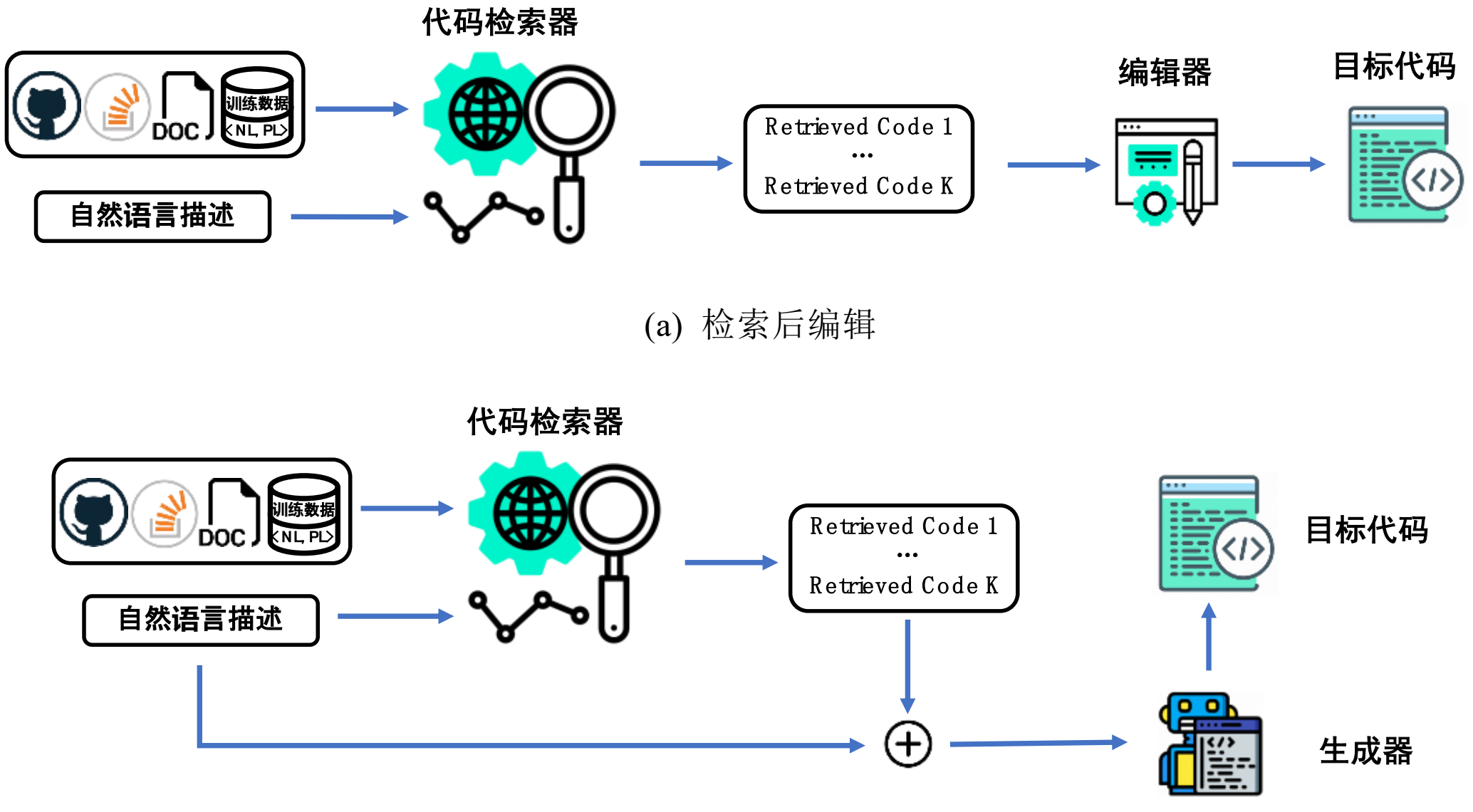

(a) 检索后编辑

(b) 检索后生成

Fig. 4 The overview of the methods with retrieval

图 4 结合检索的代码生成方法概览图

## 5 基于后处理的代码生成方法

### 5.1 简述

大规模预训练语言模型往往将代码视作普通的文本, 因此难以理解代码更深层次的语法或语义[62], 也就不能够产生具有质量和正确性保证的模型. 针对大规模预训练语言模型, 一个重要的改进方向是通过测试样例对模型进行测试, 并将生成过程及生成结果反馈给模型[63, 64, 65], 也有工作直接在模型的训练过程中利用测试样例对其进行强化, 提升模型性能[66].

通过测试样例直接对生成的代码进行改进可以视作机器执行过程的后处理, 而开发人员对生成代码进行评估可以视作人为测试代码的过程, 因此有工作利用开发人员评估意见对已有代码生成模型进行改进, 这可以被视作人工测评过程的后处理[59].

### 5.2 已有工作的分析

面对大型预训练模型可能会不能理解自然语言描述真实的意图, 同时还会产生错误代码的问题. Poesia 等人[62]设计了一种能够显著提升预训练模型代码生成可靠性的框架, 名为 SYNCHROMESH. 作者在文中提出了一种名为目标相似度调整(Target Similarity Tuning, TST)的全新语义样例选择方法, 目标是识别描述相似目标程序的自然语言来缓解概念上理解的错误; 作者定义了补全引擎(Compeletion Engines, CE), 强制要求句法合法性, 并根据语言的语义以及用户的上下文对语法限制进行编码. 当大规模语言模型按照 token 生成程序时, 限制语法解码方法(Constrainted Semantic Decoding, CSD)确保了下一个 token 是从 CE 返回的 token 集合中选择的, 这一操作解决了预训练模型产生语法错误代码的问题.

Jain 等人[63]在常规的大规模预训练语言模型的基础上加入了后处理模块(post-processing module), 对代码的语法和语义进行检查, 以确保生成的代码能够顺利通过测试样例和其他一些质量性检测. 作者在文中将大规模语言模型最常见的错误分为三类: 变量名错误(Referencing Errors, 即对于变量名的错误引用)、参数错误(Incorrect Arguments, 即对于方法内部参数值的生成错误)以及语义错误(Semantic Errors, 即生成代码形式上接近真实代码但实现功能不同). 后处理模块对代码进行修正, 也分为三类: 变量转换(Variable Transformation)、参数变换(Argument Transformation)以及抽象语法树变换(AST-to-AST Transformation). 除此之外, 作者还引入了用户反馈来改进后处理模块, 更好地基于测试样例对生成代码进行纠正.

虽然代码生成模型不断发展, 但是大多数基于深度学习的代码生成方法仍不能够保证生成代码的可编译性. Wang 等人[64]提出 COMPCODER 来提升模型生成代码的编译能力, 针对可编译的代码生成任务设计了一个全新的方法来同时训练生成器(Generator)和判别器(Discriminator). 具体而言, 作者使用强化学习得到一个预训练的代码生成器, 并同时训练一个判别器来强制让生成器纠正自己生成代码的编译错误. 模型训练可分为三个阶段, 第一个阶段是语言模型调优阶段(Language Model Fine-Tuning). 作者采用了 CodeGPT 作为生成器并在目标任务数据集上基于交叉熵损失进行调优. 第二个阶段是编译强化阶段(Compilability Reinforcement), 作者利用了强化学习, 目标是在给定输入序列的情况下, 找到能够生成具有最大化基于编译性反馈目标的输出序列的策略, 本质是进一步优化代码生成模型. 第三个阶段是编译判别阶段(Compilability Discrimination), 通过在生成器最终隐含层后添加判别器(两层带有 tanh 激活函数的 MLP), 对生成代码进行编译判别, 将模型的生成目标与判别器的判别目标结合进行训练. 这是因为作者认为具有高编译性的生成器有助于判别器的学习, 而判别优化目标可以使生成器有更强的感知能力来区分可编译代码和不可编译代码. 模型进行代码生成任务的时候, 使用生成器结合束搜索得到 top-k 个候选集, 利用判别器选择具有最高编译概率的代码作为模型最终的输出结果. 与之前常规的预训练模型相比, 使用 COMPCODER 生成的代码可编译率大幅提升.

对于大规模语言模型完成代码生成任务而言, 从生成的诸多程序中选择一个正确的解决方案是一个重要问题. 事实上, 在实际的开发场景中, 对于候选集的排序会严重影响用户的体验, 没有人会愿意看遍上百条推荐再继续选择. 评估生成代码质量与正确性的一个直接方案就是运行测试样例, 然而通过人工方式获取测试样例耗时又费力. Chen 等人[65]就聚焦于这个问题, 提出一种名为 CodeT(CODE generation with generated Test-driven dual execution greement)的方法. CodeT 利用与完成代码生成任务相同的预训练模型来自动生成测试样例, 并在生成的代码候选集上运行这些生成的测试样例, 以此考虑生成代码输出与生成测试用例的一致性, 以及当前生成代码输出与其他生成代码输出的一致性, 从而在生成代码候选集中选择最优结果.

与前面的方法相比, CodeRL[66]在利用测试用例上更进一步, 将强化学习算法引入到代码生成模型中, 试图挖掘一些在问题规范中潜在的有用信息, 解决原有大规模语言模型(作者以 CodeT5 为例)在遇到未知问题时性能较差的困境. 考虑到 CodeT5 是一个能适应多种代码任务的模型, 作者首先针对 CodeT5 中 MSP(Masked

Span Prediciton)的预训练目标进行了更改, 引入 next-token prediction 的预训练目标, 旨在弥补预训练目标任务与实际完成的代码生成任务之间的差异. 在训练过程中, 作者利用生成过程中的单元测试标识, 借助 actor-critic 框架来提升预训练语言模型的表现. critic 模型能够接收问题描述和程序来推测测试的结果, 其本质是一个分类器, 将生成的程序进行四种情况的划分({CompileError; RuntimeError; FailedTest; PassedTest}). 与此同时, 对使用基准模型产生的代码与数据集中正确代码经由测试样例产生的结果进行采样. 利用上述的分类结果和采样结果对原有模型进行优化. 在推理过程中, CodeRL 新添加了 Program Refining 和 Program Repairing, 根据相应问题的样例单元测试结果, 分别针对通过样例和失败样例对程序进行了改善和修复. 实验结果表明 CodeRL 能够作为一种模型无关的框架有效提高原有预训练模型的性能.

上面提及的方法大多通过测试过程对模型或结果加以改进, 可以视作机器完成测试从而进行后处理的过程. 事实上, 大型预训练模型在代码生成过程中采样的结果总是偏向于确定的答案(例如极短的代码片段或具有代表性的代码块). Zhang 等人[59]针对这个问题提出了 Coder-Reviewer 模型, 模型分为两个部分, Coder 是指代码生成模型, Reviewer 是评价模型, 作者试图通过 Reviewer 对 Coder 所生成的若干答案进行评估从而完成重排序. 具体来说, Coder 和 Reviewer 使用同一个模型, 借助不同的提示完成不同的功能. 给定自然语言 x, Coder 可以生成代码片段 y 并输出生成 y 的概率 P( y | x ). 在此情形下, 通过改变提示词, 给定生成的代码片段 y, Reviewer 可以得到生成代码摘要 x 的概率 P( x | y ), 最终生成代码的排序依据为两条件概率的乘积.

### 5.3 小结

本章主要对结合后处理的模型生成方法进行了简单介绍. 程序设计的目的就是解决实际问题, 测试样例与开发环境在一定程度上能够作为实际场景中问题的采样. 因此, 基于测试样例与代码实际运行环境对模型进行改进的做法一定程度上模拟了实际开发过程中所遇到的情形. 而对于遇到的问题或是错误进行修改则被称为后处理. 这种方式的处理流程类似于程序员面向 Bug 编程, 直观而有效.

目前对于测试样例的生成主要分为两类, 人工编写及模型自动生成. 对于生成测试样例的质量仍未有一个清晰的度量. 对于其他部署环境的后处理少有工作涉及.

## 6 代码生成数据集

代码生成任务用于训练, 验证和测试的数据集通常包含自然语言描述和对应的代码片段. 一个代码生成数据集一般仅包含一种高级编程语言.

目前有如下常见数据集用于训练, 验证和测试代码生成任务:

(1) HearthStone[19]: 该数据集是为卡牌游戏 HearthStone 实现的 Python 类集合, 包含 665 张不同的 HearthStone 游戏卡片, 每条数据包含一组描述卡牌信息的字段和实现了其对应功能的 Python 代码片段. 其中字段是卡牌的半结构化描述, 包含卡名, 成本, 攻击, 描述和其他属性.

(2) Magic the Gathering[19]: 该数据集是为卡牌游戏 Magic the Gathering 实现的 Java 类集合, 包含 13,297 张不同的游戏卡片. 和 HearthStone 数据集组成类似, 每条数据包含一组描述卡牌信息的字段和实现了其对应功能的 Java 代码片段.

(3) CONCODE[22]: 在 GitHub 上收集了大约 33000 个 Java 项目, 根据存储库划分 10 万个样本用于训练, 4000 个样本用于验证和测试. 基于存储库的划分使测试集中的域与训练集中的域保持分离, 因此提供了接近零样本的条件. 每个示例是由自然语言描述, 代码环境和代码段组成的元组. 其中代码环境为类中的其他成员变量和成员函数.

(4) CoNaLa[57]: 为了得到与自然语言具有细粒度对齐的代码, Yin 等人提出一种新方法. 该方法可以在 Stack Overflow 中挖掘高质量对齐数据. 实验表明该方法大大扩展了过往挖掘方法的覆盖范围和准确性. Yin 等人使用该方法创建了 CoNaLa 数据集, 其中包含 2879 个手动注释的问题及其在 Stack Overflow 上的 Python 解决方案示例, 这些示例涵盖了由具有不同意图的程序员发出的真实自然语言查询.

(5) Lyra[67]: Liang 等人认为, 在实际开发中, SQL 语句通常以字符串的形式嵌入 Python 中. 为了贴合实际应用场景, Liang 等人提出了 Turducken-Style 代码生成:为给定的自然语言注释生成具有嵌入式语言的代码, 同时发布了 Lyra 数据集. 该数据集内含自然语言注释及其对应的 Python 程序, 且该 Python 程序内含嵌入式 SQL 语句.

(6) XLCoST[68]: Zhu 等人从 GeeksForGeeks 收集了一个包含 8 种语言(C++, Java, Python, C#, JavaScript, PHP, C 和英语)的数据集 XLCoST (Cross-Lingual Code SnippeT dataset), 能够支持 10 种跨语言任务的评估. 其中, 比起之前的代码生成数据集主要针对 Python 和 Java, XLCoST 能够适应更多其他语言代码生成任务(例如 C 和 C++)的评估.

(7) MCoNaLa[69]: 为了根据英语以外的自然语言生成代码, Wang 等人提出了一个多语言数据集 McoNaLa, 用西班牙语, 日语和俄语三种语言标注了 896 个<Text, Code>对.

上述数据集大多数被划分为训练集, 验证集和测试集, 用于对模型进行训练和测试. 2020 年之前, 代码生成技术尚未成熟, 模型生成的代码大多无法执行, 因此相关研究均使用相似度作为评价指标. 在测试过程中, 通常使用 BLEU, CodeBLEU 和精确度等指标来衡量生成代码的质量, 这些指标只能衡量生成代码和参考代码的相似度. 然而, 事实上, 相似度并不能反映出代码的质量. 两段完全不同的代码可能实现了同一段自然语言描述的功能, 仅相差了一个字符的两段代码可能具有完全不同的执行结果. 仅凭借单纯的<Text, Code>对数据无法全面而完备地评判代码生成技术的好坏.

在预训练模型帮助代码生成任务取得突破后, 为了更好地体现生成代码的可执行性, 正确性, 研究人员开始通过代码的实际运行结果衡量生成代码的质量. 2020 年后测试用例数据集开始涌现. 此处列举了部分较为常用的测试用例数据集:

(1) HumanEval[44]: 2021 年 Chen 等人构建了一个人工手写的编程问题数据集 HumanEval, 内含 164 个 Python 编程问题, 每个问题包含函数头, 函数体和若干个测试用例, 平均每个问题具有 7.7 个测试用例. 这些问题评估语言理解, 算法和简单的数学, 有些与简单的编程面试问题相当.

(2) HumanEval-X[50]: 同样对 HumanEval 基准进行了扩展, 其中包含了 820 个高质量手写样本, 覆盖了 Python, C++, Java, JavaScript, Go, 可用于多种任务性能的比较(例如代码生成和代码翻译任务).

(3) APPS[70]: Hendrycks 等人从不同的开放编码网站收集编程问题, 这些问题的难度从入门到大学竞赛水平, 共包含 10000 个问题, 5000 个为训练集, 5000 个为测试集. 在测试集中, 每个问题都有多个测试用例, 平均测试用例数为 21.2 个.

(4) MBPP[71]: 内含 974 个人工编写的 Python 程序, 每个 Python 函数拥有一个简短的问题陈述和三个测试用例. 该数据集主要包含入门级编程问题.

(5) MultiPL-E[72]: 扩展了 HumanEval 基准, 使其支持 18 种以上的编程语言, 涵盖了多种编程范式.

(6) DS-1000[73]: 涉及跨越七个 Python 库的一千个数据科学问题. 从 Stack Overflow 收集用例并对其进行人工修改, 使其与原始代码不同, 从而防止模型从预训练数据中学习到相同语料造成数据泄露.

(7) AixBench[74]: 分成两个部分. 第一个部分包含 187 个功能独立, 适合单函数实现的常见 Java 任务. 每个任务包含明确的中文和英文的自然语言描述、函数签名和若干个测试用例. 第二个部分包含 160 个描述相对模糊的常见 Java 任务, 每个任务仅包含中文和英文的自然语言描述. 第二部分的生成结果需要人工评估结果的优劣.

表 3 列举了本节涉及的代码生成任务数据集及其来源.

目前代码生成领域已经出现了较丰富的数据集, 根据用途可以将其划分为两类, 一类数据集为自然语言-代码对(<Text, Code>)数据对, 主要用于对模型进行训练和测试; 另一类数据集包含编程问题和对应的测试用例, 用于评估代码的可执行能力. 早期出现的数据集多为自然语言-代码对, 利用代码相似度为作为评估. 这是由于早期的代码生成模型能力有限, 难以生成可执行代码, 在模型具有生成可执行代码的能力后, 相关研究者开始使用测试用例数据集对模型进行评估.

从数据集的规模上看，测试用例数据集的规模通常较小，这是由于测试用例通常为人工编写，开销较大.此类数据集中通常有若干编程问题，并为每个编程问题编写多个测试用例进行单元测试，从而保证评估的准确性. < Text, Code>数据集通常是从 github, Stack Overflow 等网站上爬取后整理而成，并且研究者常使用此类

Table 3 The dataset of code generation task

表 3 代码生成任务数据集

| 数据集名称 | 样本量 | 编程语言 | 输入平均长度 | 输出平均长度 | 链接 |
| --- | --- | --- | --- | --- | --- |
| Hearthstone[19] | 665 | Python | 180 | 354 | https://github.com/deepmind/card2code |
| Magic the Gathering[19] | 13297 | Java | 297 | 1082 | https://github.com/deepmind/card2code |
| CONCODE[22] | 104000 | Java | 830 | 118 | https://github.com/sriniiyer/concode |
| CoNaLa[57] | 2879 | Python/ Java | 60 | 40 | http://conala-corpus.github.io |
| Lyra[67] | 2000 | Python | - | - | https://github.com/LIANGQINGYUAN/Lyra |
| XLCoST[68] | 566752 | C++/Java/PHP/ C#/JavaScript/ Python /C | - | - | https://github.com/reddy-lab-code-research/XLCoST |
| MCoNaLa[69] | 3275 | Python/ Java | - | - | https://github.com/zorazrw/multilingual-conala |
| HumanEval[44] | 164 | Python | 451 | 181 | https://www.github.com/openai/human-eval |
| HumanEval-X[50] | 820 | Python/ C++/Go/ Java/ JavaScript | 468 | 265 | https://models.aminer.cn/codegeex/blog/ |
| APPS [70] | 10000 | Python | 1743 | 474 | https://github.com/hendrycks/apps |
| MBPP[71] | 974 | Python | 79 | 181 | https://github.com/google-research/google-research/tree/master/mbpp |
| MultiPL-E[72] | 164 | 超过 18 种 | - | - | https://github.com/nuprl/MultiPL-E |
| DS-1000[73] | 1000 | Python | 879 | 137 | https://ds1000-code-gen.github.io |
| AixBench[74] | 187+160 | Java | - | - | https://github.com/aixcoder-plugin/nl2code-dataset |

数据对模型进行训练以提升模型性能，因此< Text, Code>数据集的规模通常较大.

从表 3 的内容可以看出，目前代码生成数据集的规模虽然较大，但语言的分布存在局限性，数据集大多为单语，使用的编程语言主要为 Java 和 Python，自然语言主要为英语. 虽然目前已出现了多语数据集[69]，但为了保证模型对不同编程语言，自然语言的泛化能力，还需要相关研究者继续探索，让代码生成数据集更丰富，更多元化.

## 7 代码生成评估

为了使用统一的标准对生成代码的质量进行快速评估，研究人员使用多种指标进行评估，本章节主要使用以下自动化指标进行测试.

### 7.1 Exact Match accuracy

代码生成任务中常使用精确匹配(Exact Match)作为模型评价指标，该精确度指标表示模型生成代码与参考代码之间完全匹配的百分比.

### 7.2 BLEU

在代码生成任务中，可以将从自然语言生成代码的过程视作翻译，因此常使用 BLEU 衡量代码生成质量. BLEU[75]是一个评价机器翻译质量的指标，根据 n-gram 值计算，通过计算连续 n 个词在正确代码中的占比来衡量生成代码的相似性，其中 n-gram 是指句子中连续的 n 个 token. 代码生成任务中通常取 n 为 4，为了保证评估的准确性与公平性，计算 BLEU 值需要为 n-gram 操作引入惩罚项，BLEU 可表示为 n-gram 加权和与惩罚项的乘积.

$$BLEU = BP \cdot \exp\left(\sum_{n=1}^{N} \omega_n \log p_n\right) \tag{1}$$

$$BP = \begin{cases} 1, & c \geq r \\ e^{\{1-\frac{r}{c}\}}, & c < r \end{cases} \tag{2}$$

式(2)中 BP 表示惩罚项，c 表示模型输出代码的长度，r 表示参考代码的长度.

### 7.3 CodeBLEU

Ren 等人[76]认为，用于评价自然语言的 BLEU 指标忽略了代码的语法和语义特征，并不适合评估代码. 为了弥补这一缺陷，引入了一个新的评价指标 CodeBLEU. 它吸收了 n-gram 匹配中 BLEU 的优点，并进一步通过抽象语法树( AST )注入代码语法，通过数据流注入代码语义.

$$CodeBLEU = \alpha \cdot BLEU + \beta \cdot BLEU_{weight} + \gamma \cdot Match_{ast} + \delta \cdot Match_{df} \tag{3}$$

在 BLEU 的计算中，不同的 token 具有相同的权重，在计算 CodeBLEU 时不同的 token 具有不同的权重，$BLEU_{weight}$ 是一个加权计算的 n-gram 匹配指标; $Match_{ast}$ 是语法 AST 匹配，探索代码的语法信息; $Match_{df}$ 是语义数据流匹配，考虑了生成代码和参照代码之间的语义相似性. 加权 n-gram 匹配和句法 AST 匹配用于衡量语法正确性，语义数据流匹配用于计算逻辑正确性.

### 7.4 pass@k

BLEU 指标使用的是模糊匹配，难以衡量生成代码的可执行性. 针对这一问题，2019 年 Kulal[77]等人提出 pass@k 指标. 该指标表示被解决问题的比例，计算时需要使用编程问题数据集. 每个问题在预测时采样 k 个生成的代码样本，k 个样本中任何一个通过单元测试，则认为该问题被解决.

2021 年 Chen 等人[44]指出上述方式计算 pass@k 方差比较大，同时对该评测指标进行了改进：对每个问题，在预测时产生 n ⩾ k 个样本，统计能够通过单元测试的正确样本的数量 c ⩽ n，并且提供了修正后的 pass@k 无偏估计数值的计算函数，如图 5 所示.

```
def pass_at_k(n, c, k):
    """
    :param n: total number of samples
    :param c: number of samples to pass
    :param k: k in pass@k
    """
    if n - c < k:
        return 1.0
    return 1.0 - np.prod(1.0 - k / np.arrange(n - c + 1, n + 1))
```

Fig. 5 A numerically stable script for calculating unbiased estimates for pass@k[44]

图 5 用于计算 pass@k 的无偏估计的数值稳定脚本[44]

## 8 总结与展望

### 8.1 总结

本文对高级程序代码生成任务目前国内外最新进展进行了比较详尽的阐述与总结. 我们将当前的智能化代码生成技术分为了三类, 第一类是基于代码特征的代码生成方法, 第二类是结合检索的代码生成方法, 第三类是结合后处理的代码生成方法.

基于代码特征的代码生成方法又可以根据模型训练的范式分为基于监督学习的代码生成方法和基于预训练的代码生成方法. 前者使用了传统的编码器-解码器架构以及 RNN 与 CNN 模型, 并通过加入结构化信息, 对解码器的改进等操作, 在相关数据集上取得了不错的效果. 后者则是从自然语言处理领域的预训练大规模模型得到灵感, 使用软件工程领域相关代码数据和目标函数对大型语言模型进行预训练与调优, 从而大幅提升模型的性能.

结合检索的代码生成方法和结合后处理的代码生成方法可视作对第一类方法部分模块的改进, 试图提升代码生成模型的性能.

对于结合检索的模型而言, 检索的方式常常用于对已有的代码生成模型进行增强, 可视作即插即用的组件式模块. 一方面, 检索到的内容更多作为输入数据的增强帮助原有模型更好地完成生成任务[54, 58]; 另一方面, 检索能够作为桥梁以便更加直接利用外部的文本和源码信息作为原有模型的补充[56].

结合后处理的模型通过实际运行过程借助测试样例和开发环境对生成的代码进行处理, 以提升其实际运行表现, 一个趋势是测试样例的产生由人为编写[63]转变为模型产生[65]. 不同的模型对于不同的测试样例的用法不尽相同, 但是核心思想和真实场景下编程相似, 即通过一些可能的错误来使得生成的代码更加健壮.

除此之外, 本文还概括介绍了智能化代码生成任务中常用的数据集与评价指标, 以方便该领域后续研究能够进行合理的实验设计.

### 8.2 展望

虽然目前的代码生成方法在相关的数据集上已经取得了较好的结果, 但是相关技术仍然拥有许多挑战值得关注.

**(1) 自然语言描述与代码的对齐问题**

自然语言与代码在描述信息的角度与方法上有着巨大差异, 传统上直接利用 token 训练得到的序列到序列模型很难去理解并提取自然语言中的功能表述并翻译成对应的代码. 所以将代码表示为树来有效引入代码的结构信息成为提升模型性能的重要举措, 但这并不是一项容易的任务. 语法树中的节点数量通常大大超过其对应自然语言描述的长度, 这样的不对称性会增加代码生成任务的难度. 因此, 如何将自然语言描述与代码进行对齐对于传统的序列到序列模型是一个极大的挑战.

另一方面, 模型对于输入自然语言的理解对于代码生成任务的效果也格外重要. 自然语言描述多种多样, 对于同样的意思可能有多种表达的方式. 对输入的多种自然语言进行重构, 为生成的代码提供更加清晰明了的需求, 可能是一项解决方法.

除此之外, 之前有工作[27, 28, 58, 59]将代码生成与代码摘要两个任务作为对偶任务对同一模型进行训练取得了不错的效果, 尤其是 CodeT5 添加对偶任务后与 CoTexT 相比在代码生成任务上的性能有显著提升. 因此, 如何将代码生成与代码摘要生成两个任务更好的结合在一起提升性能仍需要被探索.

**(2) 大规模预训练模型的语法正确性问题**

基于 AST 和语法的模型能够产生语法正确的代码, 但是一定程度上忽略了语义的正确性[31]; 预训练模型大多能保证语义的信息, 但是不能保证语法的正确性[62].

由于自然语言和代码之间本身就存在的语义差距, 使得代码生成模型没有足够的规则约束代码的生成过程, 造成了生成目标代码时庞大的解空间. 这样的问题在预训练模型上表现尤为明显, 虽然在大量代码库上进行预训练所得的大规模模型可以生成语义相关的代码片段, 但是生成的代码很容易具有语法错误且存在可

读性较低的问题. 在解空间中找到最优解或较优解成为了一项难题. 如何学习并利用好代码中有效的结构信息, 填补自然语言与代码之间的差异, 从而对解码过程进行约束, 提高生成代码的可靠性仍是一个待探究的方向

现有的预训练模型大多都是继承于自然语言处理领域的相关技术, 代码领域相关研究一直沿着自然语言处理领域的道路前进. 因此, 自然语言处理中的语法正确性问题在代码生成任务中也普遍存在. 但有研究表明, 自然语言处理中的嵌入方法可能在代码领域中不是最优的, 脱离自然语言处理思维模式可能是一种优化的思路[78].

**(3) 大规模预训练模型能力的挖掘**

虽然近年来应用于代码生成任务的大规模语言模型层出不穷, 但是已有许多论文[17, 18]表明随着训练的继续进行, 模型的性能还在持续增长. 一些预训练模型还根据模型的大小将模型分为 small, base 和 large 来适应不同的硬件条件不同需求下模型的应用. Wang 等人[79]将自然语言处理中的提示学习(prompt learning)应用到代码相关的分类任务上, 取得的性能的提升. 2022 年初随着思维链(chain-of-thought)[80]的提出, 大规模预训练模型的知识被进一步挖掘. 鉴于 ChatGPT 强大的性能, 以及人类程序员在代码编写过程中的步骤化和模块化的考量, 将提示学习引入代码生成任务, 更深一步挖掘已有大规模预训练模型的性能在未来可能会成为一个重要的研究方向.

除此之外, 还有相关研究[81, 82]发现了将 Stack Overflow 中的问题和代码答案作为双模态数据用于模型调优有助于模型性能的提升以及错误代码生成的抑制, 证明了这样的外部数据能够成为挖掘大规模模型性能的有效工具.

**(4) 模型适应多种不同编程语言的能力**

目前虽然能够完成代码生成任务的模型有很多, 但是对于高级程序语言而言, 大多围绕 Java 和 Python 展开. 这一方面是由于数据集的原因, 作为当今最为主流的两大语言, Java 和 Python 在数据量和数据质量上都拥有得天独厚的优势. 另一方面是结构的问题, Java 和 Python 都是面向对象的编程语言, 与其它编程语言(比如 C 语言)相比, 因为完备的库与 API 接口为相关功能的实现提供支撑, 所以他们可以完成具有更为简洁直观的编写方式, 在一定程度上与自然语言相匹配. 最近, 有一些工作创建了多种语言的代码库[10, 67, 68], 如何合理利用这样的代码库让模型能够适应多种不同编程语言, 或许是提升模型性能的一种方式.

**(5) 大型预训练模型压缩的问题**

目前对于代码生成类工具的普遍解决方案是将本地的代码和软件开发人员的相关需求传送给服务器端, 由服务器端完成计算并将结果回传给本地. 由于服务器端的模型需要实时的代码和需求说明才能够进行计算并反馈结果, 因此代码泄露成为这个过程中不可忽视的一个隐藏问题. 为解决代码泄漏的问题, 将高性能模型移植到本地或其他受限资源的平台上逐渐成为一项巨大的挑战.

此外, 代码生成任务的训练和推理过程都对服务器端产生了巨大的压力, 伴随着模型的不断增大, 如果将训练及推理代价与实际的计算资源进行权衡, 使得模型性能维持相对较高水平的情况下节省资源, 缩短代码生成时间, 为用户带来更好的体验成为了第二项重大的挑战.

以上两项挑战的解决可以归结于大型模型的压缩, 一方面可以将模型的规模变小, 使其能够部署在一些较低资源的平台上, 节省计算资源; 另一方面可以提升模型的推理速度, 带来代码生成工具的体验的提升.

**(6) 人工评估代码生成方法性能**

代码生成技术的意义在于辅助软件开发人员在实际开发过程中进行编码, 要想评估代码生成技术的实际价值, 就需要对其进行人工评估, 人为地评判该技术在工程实践中所发挥的作用. 但目前人工评估存在着评判标准模糊, 评价指标不统一, 难以大规模快速评估的问题, 同时考虑到人工评估的代价巨大, 因此目前仅有少量研究进行人工评估[83, 84], 但如果想要现在的优秀模型能够落地成为普惠广大用户的代码生成工具, 人工评估方法的确定, 实际测试和广泛讨论必不可少.

除此之外, ChatGPT 被视作是与人类高度对齐的一个模型, 并被广泛用于类 GPT3.5 模型的评估[85, 86]. 如

何更好地使用 ChatGPT 对生成的代码进行打分, 并从更高的维度对生成的代码进行评价尚未被探索. 而这种方式很可能是类人工自动化评估代码的未来之路.